\documentclass[aps,prd,twocolumn,showkeys,superscriptaddress]{revtex4-2}
\usepackage{amsmath,amssymb}
\usepackage{graphicx}

\newcommand{\bastar}{\begin{eqnarray*}}
\newcommand{\eastar}{\end{eqnarray*}}
\newskip\humongous \humongous=0pt plus 1000pt minus 1000pt

\newif\ifdtup

\relax
\newcommand{\W}{{\vec W}}
\newcommand{\n}{{\hat n}}

\newcommand{\hD}{{\hat D}}
\newcommand{\bea}{\begin{eqnarray}}
\newcommand{\eea}{\end{eqnarray}}
\newcommand{\pd}{\partial}
\newcommand{\A}{{\vec A}}
\newcommand{\hF}{{\hat F}}
\newcommand{\hA}{{\hat A}}
\newcommand{\F}{{\vec F}}

\newcommand{\nn}{\nonumber}

\newcommand{\mn}{{\mu\nu}}
\begin{document}
\title{Electroweak Monopole-Antimonopole Pair Production at LHC}
\bigskip

\author{Petr Benes}
\email{petr.benes@utef.cvut.cz}
\affiliation{Institute of Experimental and Applied Physics, \\ Czech Technical University in Prague, Husova~240/5, 110~00 Prague~1, Czech Republic}
\author{Filip Blascke}
\email{filip.blaschke@fpf.slu.cz}
\affiliation{Research Centre for Theoretical Physics and Astrophysics, Institute of Physics, \\ Silesian University in Opava, Bezru\v{c}ovo n\'{a}m\v{e}st\'{\i}~1150/13, 746~01 Opava, Czech Republic}
\author{Y. M. Cho}
\email{ymcho0416@gmail.com}
\affiliation{School of Physics and Astronomy, 
Seoul National University, Seoul 08826, Korea}
\affiliation{Center for Quantum Spacetime, Sogang University, Seoul 04107, Korea}

\begin{abstract}
It is argued that the monopole production at LHC crucially depends on the monopole production mechanism. We show that, if the monopole production mechanism 
at LHC becomesthe thermal fluctuation of the Higgs vacuum as the early universe did, it is practically impossible for LHC to produce the monopole. This is because the temperature of the p-p fireball is simply too low to generate the thermal fluctuation necessary 
for the monopole production. But if the monopole production mechanism becomes the Drell-Yan and/or Schwinger mechanism, the 14 TeV LHC could produce 
the monopole only when the mass is less than 7 TeV. 
To circumbent this energy constraint, we propose 
a new monopole production mechanism at LHC, the monopole production in the form of the monopolium
bound state. We show that LHC could produce 
the monopole in the form of the monopolium of mass 
less than 14 TeV, even when the monopole mass becomes considerably larger than 11 TeV. In this case we can prove the existence of the electroweak monopole 
at present LHC by detecting the decay modes of 
the monopolium. This implies that the monopole production in the form of the monopolium could be 
the most probable way to detect the electroweak monopole at LHC. We discuss the physical implications of our result.         
\end{abstract}
.

\maketitle

\section{Introduction}

With the advent of the Dirac's monopole 
the magnetic monopole has become an obsession 
in physics, experimentally as well as 
theoretically \cite{dirac,cab}. After the Dirac monopole we have had the Wu-Yang monopole, 
the 'tHooft-Polyakov monopole, and the grand unification monopole \cite{wu,thooft,dokos}. But 
the electroweak (``Cho-Maison") monopole stands 
out as the most realistic monopole that could 
exist in nature and could actually be 
detected \cite{plb97,yang}. 

Indeed the Dirac monopole in electrodynamics should transform to the electroweak monopole after 
the unification of the electromagnetic and weak interactions, and the Wu-Yang monopole in QCD is 
supposed to make the monopole condensation to confine 
the color. Moreover, the 'tHooft-Polyakov monopole 
exists only in an hypothetical theory, and the grand unification monopole which could have been amply produced at the grand unification scale in the early universe probably has become completely irrelevant 
at present universe after the inflation. 

This makes the experimental confirmation of 
the electroweak monopole one of the most urgent 
issues in the standard model after the discovery 
of the Higgs particle \cite{pta19,iguro,gould}. In fact 
the detection of this monopole, not the Higgs particle, should be regarded as the final (and topological) test of the standard model. For this reason the MoEDAL and ATLAS detectors at LHC are actively searching for 
the monopole \cite{medal1,medal2,medal3,atlas}. 

To detect the electroweak monopole at LHC, we need 
to remember the basic facts about the monopole \cite{plb97,yang,pta19,epjc15,ellis,bb,epjc20}. 
First, this is the monopole which exists within 
(not beyond) the standard model as the electroweak generalization of the Dirac monopole, which can be viewed as a hybrid between Dirac and 'tHooft-Polyakov monopoles. Second, the magnetic charge of the monopole is not $2\pi/e$ but $4\pi/e$, twice that of 
the Dirac monopole. This is because the period of 
the electromagnetic U(1) subgroup of the standard 
model becomes $4\pi$. Third, the mass of the monopole $M$ is of the order of 10 TeV. This is because 
the mass basically comes from the same Higgs 
mechanism which makes the W boson massive, except 
that here the magnetic coupling makes the monopole 
mass $1/\alpha$ times heavier than the W boson mass, 
around 11.0 TeV \cite{zel}. Despite this, the size of 
the monopole is set by the W boson mass, because 
the monopole solution has the W (and Higgs) boson   
dressing which fixes the size by the W boson mass. 

Because of these distinctive features, MoEDAL detector 
could identify the monopole without much difficulty, 
if LHC could produce it. However, the 14 TeV LHC may 
have no chance to produce the monopole if the mass 
becomes larger than 7 TeV. This is problematic, because 
the monopole mass could turn out to be bigger than this, around 11 TeV. If this is so, we may have little chance to detect the monopole at LHC, and may have to try to detect the remnant monopoles at present universe produced in the early universe \cite{pta19,cko}.

However, two observations could make the monopole 
production at LHC possible even when the mass becomes heavier than 7 TeV. First, the monopole has to be 
created in pairs at LHC. This implies that at the initial stage, the monopoles could appear in the form of the atomic monopolium states made of the monopole and anti-monopole pair. In this case there is 
the possibility that the binding energy could reduce the bound state of monopole-antimonopole pair below 
14 TeV, even if the monopole mass becomes bigger 
than 7 TeV. 

The second point is related to the electroweak 
monopole production mechanism at LHC. At present 
there are three contending monopole production mechanisms. Two popular ones in the high energy 
physics community are the Drell-Yan (and two photon fusion) process \cite{medal1,medal2} and the Schwinger mechanism \cite{medal3,schw,ho}. The third one is 
the topological monopole production mechanism, in 
which the thermal fluctuation of the Higgs vacuum produces the monopole as the early universe 
did \cite{pta19,cko}. LHC could produce the monopole 
with this mechanism because the fireball of the p-p collision could reproduce the early universe. 

A unique feature of this topological thermal fluctuation mechanism is that in this production 
the monopole mass depends on the temperature, 
so that at the creation the initial monopole 
mass could be smaller than the zero temperature 
mass \cite{pta19,cko}. In this case there is 
a possibility that LHC could produce the baby electroweak monopole even when the zero temperature mass becomes heavier than 7 TeV. 

{\it The purpose of this paper is to discuss how 
LHC can accommodate the above ideas and produce 
the electroweak monopoles even when the monopole 
mass exceeds 7 TeV. We confirm that, if the monopole production mechanism at LHC is the Drell-Yan process (and/or Schwinger mechanism), the present 14 TeV LHC could produce the monopole only if the mass is less than 7 TeV. However, if the monopole production mechanism at LHC becomes the thermal fluctuation of 
the Higgs vacuum, it has practically no chance to 
produce the monopole (even with the future FCC). 
This is because the temperature of the p-p fireball 
at LHC simply becomes too low to produce the monopole thermally. We propose a new monopole production mechanism at LHC, the monopole production in the form of the monopolium, which could allow LHC to produce 
the electroweak monopole even when the monopole mass 
becomes larger than 7 TeV. In specific, we show that LHC could produce the monopolium bound states of mass around 5.7 TeV for any reasonable monopole mass, as 
far as the mass does not exceed 29.8 TeV. This is because the binding energy of the monopoliun could reduce the monopolium mass by 16.3 TeV. This makes 
the monopolium bound state a most probable signal 
for the electroweak monopole production at LHC.}

The paper is organized as follows. In Section II we 
discuss two electroweak monopole production mechanisms 
at LHC, Drell-Yan process and Schwinger mechanism. 
In Section III we review the monopole production 
in the electroweak phase transition in the early universe. In Section IV we discuss if LHC could produce the monopole in a similar manner, and argue that this is practically impossible because the temperature of the p-p fireball is too low. In Section V we discuss the naive Bohr model of thelectroweak monopolium and show that this model is unrealistic to describe an electroweak monopolium. In Section VI we discuss 
a more realistic electroweak monopole production mechanism, the monopole production via monopolium, 
and argue that such monopolium bound state could be produced at LHC. In Section VII we discuss 
the physical impications of our results.   

\section{Electroweak Monopole Production Mechanism at LHC: Drell-Yan Process and Schwinger Mechanism} 

Consider the (bosonic sector of the) Weinberg-Salam 
Lagrangian,
\begin{gather}
{\cal L}_{WS} = -|{\cal D}_\mu \phi|^2
-\frac{\lambda}{2}\big(\phi^\dagger \phi
-\frac{\mu^2}{\lambda} \big)^2
-\frac{1}{4} \vec F_\mn^2
-\frac{1}{4} G_\mn^2, \nn\\
{\cal D}_\mu \phi = \big(\pd_\mu
-i\frac{g}{2} \vec \tau \cdot \A_\mu 
- i\frac{g'}{2}B_\mu \big) \phi,
\label{wslag}
\end{gather}
where $\phi$ is the Higgs doublet, $\vec F_\mn$ and $G_\mn$ with potentials $\A_\mu$ and $B_\mu$ are the gauge fields 
of $SU(2)$ and $U(1)_Y$, $\cal D_\mu$ is the covariant derivative, and $g$ and $g'$ are the corresponding coupling 
constants. Expressing $\phi$ with the Higgs field $\rho$ 
and unit doublet $\xi$ by
\begin{gather}
\phi = \dfrac{1}{\sqrt{2}}\rho~\xi,
~~~(\xi^\dagger \xi = 1),
\end{gather}
we have
\begin{gather}
{\cal L}_{WS}=-\frac{1}{2} (\pd_\mu \rho)^2
- \frac{\rho^2}{2} |{\cal D}_\mu \xi |^2
-\frac{\lambda}{8}\big(\rho^2-\rho_0^2 \big)^2 \nn\\
-\frac14 \vec F_\mn^2 -\frac14 G_\mn^2,
\end{gather}
where $\rho_0=\sqrt{2\mu^2/\lambda}$ is the vacuum 
expectation value of the Higgs field.

We can express (\ref{wslag}) in terms of the physical 
fields gauge independently \cite{pta19}. With the Abelian decomposition of $\A_\mu$ 
\begin{gather}
\A_\mu = \hA_\mu +\W_\mu,  \nn\\
\hA_\mu= A_\mu \n -\frac{1}{g} \n \times \pd_\mu \n, 
~~~\n =-\xi^\dag \vec \tau \xi,   \nn\\
\F_\mn=\hF_\mn + \hD _\mu \W_\nu - \hD_\nu
\W_\mu + g\W_\mu \times \W_\nu,   \nn\\
\hD_\mu=\pd_\mu+g \hA_\mu \times,
\end{gather}
we have
\begin{gather}
\F_\mn^2 = {F'}_\mn^2 + 2 |D'_\mu W_\nu-D'_\nu W_\mu|^2 
-4ig F'_\mn W_\mu^*W_\nu   \nn\\
-g^2 (W_\mu^* W_\nu - W_\nu^* W_\mu)^2,  \nn\\
F'_\mn= \pd_\mu A'_\nu-\pd_\nu A'_\mu,  
~~A'_\mu =A_\mu + C_\mu,
~~C_\mu =-\frac{2i}{g} \xi^\dag \pd_\mu \xi  \nn\\
D'_\mu=\pd_\mu + ig A'_\mu,
~~~W_\mu =\frac{W^1_\mu + i W^2_\mu}{\sqrt{2}}. 
\label{qcdadec}
\end{gather}
With this we can define $A^{\rm (em)}$ and $Z_\mu$ by
\begin{gather}
\left( \begin{array}{cc} A_\mu^{\rm (em)} \\  Z_{\mu}
\end{array} \right)
=\frac{1}{\sqrt{g^2 + g'^2}} 
\left(\begin{array}{cc} g & g' \\
-g' & g \end{array} \right)
\left( \begin{array}{cc} B_{\mu} \\ A'_{\mu}
\end{array} \right)  \nn\\
= \left(\begin{array}{cc}
\cos\theta_{\rm w} & \sin\theta_{\rm w} \\
-\sin\theta_{\rm w} & \cos\theta_{\rm w}
\end{array} \right)
\left(\begin{array}{cc} B_{\mu} \\ A_\mu'
\end{array} \right), 
\label{mix}
\end{gather}
and have the identity
\begin{gather}
|{\cal D}_\mu \xi|^2 =\frac14 Z_\mu^2
+\frac{g^2}{2} W_\mu^* W_\mu.  
\end{gather}
Notice that here we have defined $A_\mu^{(em)}$ and $Z_\mu$ gauge independently, without any reference to the unitary gauge.

\begin{figure}[t]
\includegraphics[height=4cm, width=7cm]{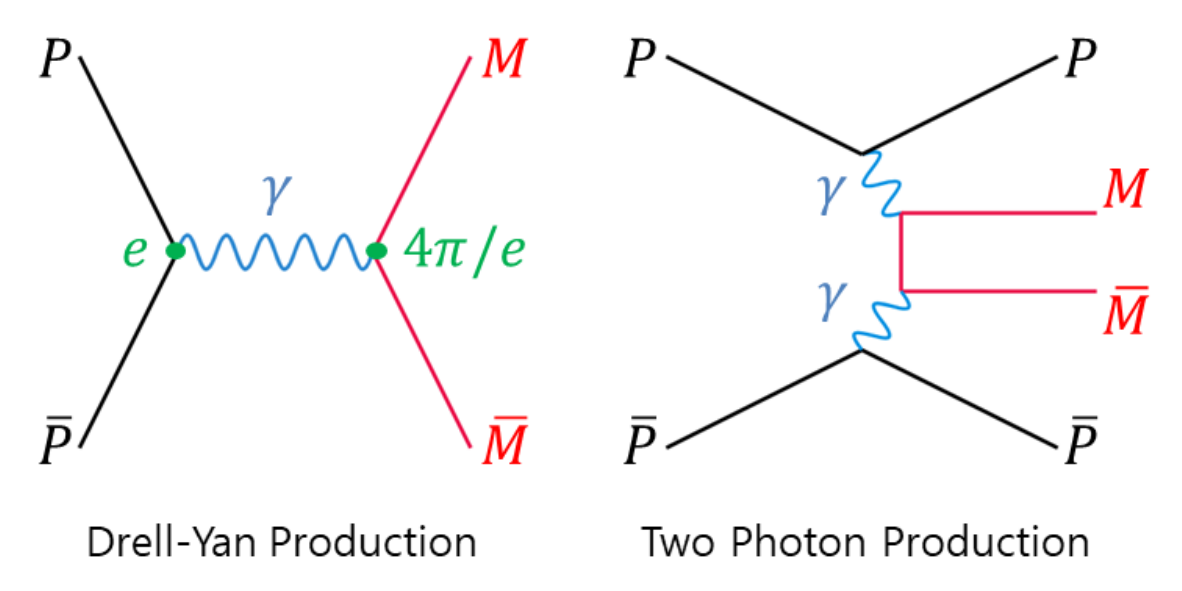}
\caption{\label{dyp} The Feynman diagrams of 
the popular monopole production mechanism at LHC 
given by Drell-Yan and two-photon fusion process. 
Here the proton pair could also be interpreted 
as the quark pair.}
\end{figure}

From this we can express the Weinberg-Salam Lagrangian in terms of the physical fields,
\begin{gather}
{\cal L}_{WS} = -\frac{1}{2}(\pd_\mu \rho)^2 
-\frac{\lambda}{8}\big(\rho^2-\rho_0^2 \big)^2 \nn\\
-\frac14 {F_\mn^{\rm (em)}}^2 
-\frac14 Z_\mn^2-\frac{g^2}{4}\rho^2 W_\mu^*W_\mu
-\frac{g^2+g'^2}{8} \rho^2 Z_\mu^2 \nn\\
-\frac12 |(D_\mu^{\rm (em)} W_\nu -D_\nu^{\rm (em)} W_\mu)
+ ie \frac{g}{g'} (Z_\mu W_\nu - Z_\nu W_\mu)|^2  \nn\\
+ie F_\mn^{\rm (em)} W_\mu^* W_\nu
+ie \frac{g}{g'}  Z_\mn W_\mu^* W_\nu \nn\\
+ \frac{g^2}{4}(W_\mu^* W_\nu - W_\nu^* W_\mu)^2,
\label{wsadec}
\end{gather}
where $D_\mu^{\rm (em)}=\pd_\mu+ieA_\mu^{\rm (em)}$
and $e$ is the electric charge
\begin{gather}
e=\frac{gg'}{\sqrt{g^2+g'^2}}=g\sin\theta_{\rm w}
=g'\cos\theta_{\rm w}.
\label{e}
\end{gather}
We emphasize that this is not the Weinberg-Salam Lagrangian in the unitary gauge. This is the gauge independent abelianization of the Weinberg-Salam Lagrangian. Notice 
that here the Higgs doublet disappears completely, and 
the Higgs, W, and Z bosons acquire the mass $M_H =\sqrt{\lambda} \rho_0$, $M_W =g \rho_0/2$, $M_Z =\sqrt{g^2+g'^2} \rho_0/2$, without any spontaneous 
symmetry breaking.    

\begin{figure}
\includegraphics[height=4cm, width=6cm]{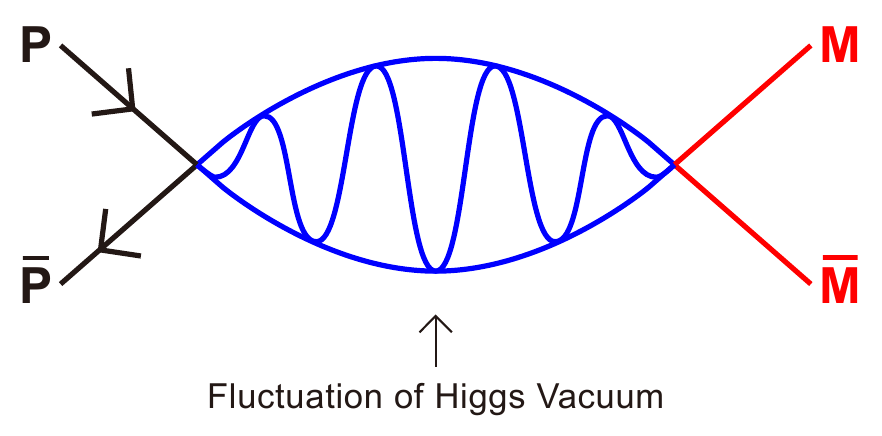}
\caption{\label{topro} The topological monopole production mechanism in the early universe or at LHC induced by the thermal fluctuation of the Higgs vacuum.}
\end{figure}

The Lagrangian has the monopole topology $\pi_2(S^2)$
that comes from the unit doublet $\xi$ which could 
be identified as a $CP^1$ field. It has monopole solutions, the naked Cho-Maison monopole and the Cho-Maison monopole dressed by the W and Higgs bosons \cite{plb97,yang}. Although the Cho-Maison monopole has infinite energy classically, one could predict the mass to be around 
4 to 11 TeV. Intuitively we could argue the mass to be $1/\alpha$ times bigger than the W boson mass, around 
11 TeV. This is because the monopole mass comes from 
the same Higgs mechanism that makes the W boson massive, except that here the gauge coupling is magnetic 
(i.e., $4\pi/e$) \cite{zel}. We could backup this 
arguement regularizing the Cho-Maison monopole, 
and estimate the mass to be around 4 to 
10 TeV \cite{epjc15,ellis,bb,pta19,epjc20}. But 
in the following we will assume for simplicity 
that the electroweak monopole mass is $M_W/\alpha$, 
around 11.0 TeV.

To understand the electroweak monopole production 
at LHC we have to know the monopole production mechanism at LHC. In the Drell-Yan process 
the monopole-antimonopole are produced in pairs 
by the electromagnetic interaction through 
the photon \cite{medal1,medal2}. But in the Schwinger mechanism a strong magnetic field is supposed to 
create the monopole-antimonopole pair, just like 
a strong electric background creates 
the electron-positron pairs in the non-perturbative 
QED \cite{medal3,schw}. The justification for this comes from the asumption that a theory of monopole 
must be symmetric under the dual transformation 
$(\vec E, \vec H)$ to $(\vec H, -\vec E)$ together 
with $(e,g)$ to $(g,-e)$. The Drell-Yan process 
is shown graphically in Fig. \ref{dyp}.

But in the topological monopole production, the thermal fluctuation of the Higgs vacuum induces the change of topology and produce the monopole, which is exactly 
the monopole production mechanism in the early 
universe \cite{pta19}. In this picture the monopoles 
are produced after the phase transition and stops at 
the Ginzburg temperature. The topological monopole production mecanism is shown graphically in 
Fig. \ref{topro} for comparison.

Notice that, in the Drell-Yan process the monopole 
pair is produced via the photon solely by 
the electromagnetic process. But the Weinberg-Salam Lagrangian has no interaction which can be described 
by the above Feynman diagram perturbatively. 
Similarly, for the Schwinger mechanism the one-loop effective action of the Weinberg-Salam theory has 
no indication of the monopole pair production in 
strong magnetic background. So these monopole production mechanisms are possible beyond (not within) the standard model. More importantly, these mechanisms 
do not take into account the change of topology 
necessary to produce the monopole, so that 
the topological nature of the monopole is completely neglected in these mechanisms.   

In comparison the topology plays an essential role in 
the thermal production of the monopole, because here 
the thermal fluctuation of the Higgs vacuum is precisely what we need to induce the change of topology. As a result the monopole production in this mechanism is not controlled by any fundamental interaction or fundamental constant. Perhaps more importantly, in this thermal fluctuation the monopole mass depends on the temperature at creation
because the vacuum value of the Higgs field which 
determines the monopole mass does. This means that 
the monopole mass at creation could be considerably 
smaller than the monopole mass at zero temperature. In 
this case we have to distinguish the finite temperature monopole mass from the zero temperature monopole mass. 
So, from now on we call the initial monopole mass (the monopole mass at birth) as the ``baby" (or ``infant") monopole mass and the zero temperature mass as the ``adolescent" monopole mass. 

It should be emphasized that the above monopole production mechanisms at LHC are theoretical 
(i.e., logical) but not realistic possibilities. 
For example, the Schwinger mechanism is certainly 
a logical possibility. However, it is not clear 
at all if (and how) the p-p fireball at LHC could create a strong magnetic background which can 
actually produce the monopole-antimonopole pair.  

If so, what is the monopole production mechanism at LHC? We do not know yet. On the other hand, it has often been claimed that LHC could reproduce the early universe. If this is true, the monopole production mechanism at LHC could be thermal, just like the electroweak monopole production in the early universe. To discuss if this is true, we have to understand 
the monopole production mechanism in the early universe. 

\section{Electroweak Monopole Production in the Early Universe: A Review}

It has generally been believed that the monopole production in the early universe critically depends 
on the type of the phase transition. In the first 
order phase transition the vacuum bubble collisions 
in the unstable vacuum are supposed to create the monopoles through the quantum tunneling to the stable vacuum during the phase transition, so that the monopole production is supposed to be suppressed
exponentially by the vacuum tunneling \cite{guth}. 
On the other hand the monopole production in 
the second order phase transition is supposed to be described by the Kibble-Zurek mechanism which has 
no such exponential suppression \cite{kibb,zurek}.
 
We emphasize, however, that this popular view may 
have a critical defect \cite{pta19}. This is because 
in the topological monopole production mechanism 
the thermal fluctuation of the Higgs vacuum which provides the seed of the monopoles continues 
untill the temperature drops to the Ginzburg 
temperature \cite{gin}. This means that, even 
in the first order phase transition we can have 
the monopole production after the vacuum bubble tunneling without the exponential suppression, 
if the Ginzburg temperature becomes less than 
the critical temperature. In this case the monopole production in the first order phase transition 
becomes qualitatively the same as in the second 
order phase transition. This tells that the popular exponential suppression of the monopole production 
in the first order phase transition is only 
half of the full story which could be totally misleading.

In the second order phase transition the thermal fluctuation of the Higgs vacuum could also modify 
the Kibble-Zurek mechanism considerably, because it provides more time for the monopole production as 
the thermal fluctuation could continue long after 
the phase transition. This tells that what is important in the monopole production in the early universe is 
the Ginzburg temperature, not the type 
of the phase transition.

To amplify this point we start from the temperature 
dependent effective action of the standard model 
which describes the electroweak phase transition 
and the monopole production in the early 
universe \cite{pta19,kriz,and,cko}
\begin{gather}
V_{eff}(\rho) =V_0(\rho) -\frac{C_1}{12\pi} \rho^3~T
+\frac{C_2}{2} \rho^2~T^2 
-\frac{\pi^2}{90} N_* T^4  \nn\\
+\delta V_T,  \nn\\
V_0(\rho)=\frac{\lambda}{8}(\rho^2-\rho_0^2)^2 ,  \nn\\
C_1=\frac{6 M_W^3 + 3 M_Z^3}{\rho_0^3}\simeq 0.36,   \nn\\
C_2=\frac{4M_W^2 +2 M_Z^2 +M_H^2+4m_t^2}{8\rho_0^2} 
\simeq 0.37,   
\label{epot}
\end{gather}
where $V_0$ (with $\lambda \simeq 0.26$ and 
$\rho_0 \simeq 246~{\rm GeV}$) is the zero-temperature 
potential, $C_1$ and $C_2$ terms are the loop contributions from the gauge bosons, Higgs field, 
and heavy fermions, $N_*$ is the total number of distinct helicity states of the particles with mass smaller than $T$ (counting fermions with the factor 7/8), $m_t$ is the top quark mass, and $\delta V_T$ 
is the slow-varying logarithmic corrections and 
the lighter quark contributions which we will 
neglect from now on.

The potential has three local extrema at 
\begin{gather}
\rho_s=0,   \nn\\
\rho_{\pm}(T)=\Big\{\frac{C_1}{4\pi \lambda}
\pm \sqrt{\Big(\frac{C_1}{4\pi \lambda} \Big)^2
+\frac{\rho_0^2}{T^2} 
-\frac{2C_2}{\lambda}} \Big\}~T^.
\label{rext}
\end{gather}
The first extremum $\rho_s=0$ represents the Higgs vacuum of the symmetric (unbroken) phase, the second one $\rho_-(T)$ represents the local maximum, and 
the third one $\rho_+(T)$ represent the local minimum Higgs vacuum of the broken phase. But the two extrema $\rho_{\pm}$ appear only when $T$ becomes smaller 
than $T_1$
\begin{gather}
T_1 =\frac{4 \pi \lambda}{\sqrt{32\pi^2\lambda C_2-C_1^2}}
~\rho_0 \simeq 146.74~\text{GeV}.
\end{gather}
So above this temperature only $\rho_s=0$ becomes 
the true vacuum of the effective potential, and 
the electroweak symmetry remains unbroken. 

\begin{figure}
\includegraphics[height=5.5cm, width=7cm]{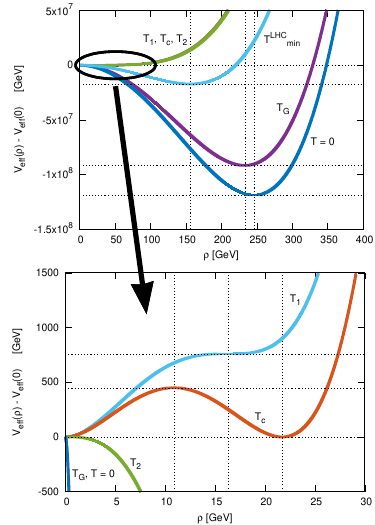}
\caption{\label{tpot} The effective potential (\ref{epot}) at different temperatures. Notice that the potential at $T_1,~T_c,~T_2$ are almost indistinguishable. Here the unit of $V_{eff}$ is chosen to be $V_0=(\lambda/8) \rho_0^4=1$.}
\end{figure}

At $T=T_1$ we have 
\begin{gather}
\rho_-=\rho_+=(C_1/4\pi \lambda)~T_1 \simeq 16.3~{\rm GeV},
\end{gather} 
but as temperature cools down below $T_1$ we have two 
local minima at $\rho_s$ and $\rho_+$ with 
$V_{eff}(0)< V_{eff}(\rho_+)$, until $T$ reaches 
the critical temperature $T_c$ where $V_{eff}(0)$ 
becomes equal to $V_{eff}(\rho_+)$,
\begin{gather}
T_c= \sqrt{\frac{18}{36\pi^2 \lambda C_2- C_1^2}} 
~\pi \lambda \rho_0  \simeq 146.70 ~{\rm GeV},   \nn\\
\rho_+(T_c)=\frac{C_1}{3\pi \lambda}~T_c
\simeq 21.7~{\rm GeV}.
\label{ctemp}
\end{gather}
So $\rho_s=0$ remains the minimum of the effective 
potential for $T>T_c$. Notice that $\rho_+(T_c)/\rho_0 \simeq 0.09$. 

At $T_c$ the new vacuum bubbles start to nucleate 
at $\rho_s=0$, which tunnels to the stable vacuum $\rho_+$ after $T_c$. Below this critical temperature $\rho_+$ becomes the true minimum of the effective potential, but $\rho_s=0$ remains a local minimum 
till the temperature reaches $T_2$,
\begin{gather}
T_2=\sqrt{\frac{\lambda}{2C_2}}~\rho_0 
\simeq 146.42~\text{GeV},   \nn\\
\rho_+(T_2) = \frac{C_1}{2\pi \lambda}~T_2 
\simeq 32.5~\text{GeV}.
\end{gather} 
From this point $\rho_+$ becomes the only (true) minimum, which approaches to the well-known Higgs vacuum $\rho_0$ at zero temperature. The effective potential (\ref{epot}) is shown in Fig. \ref{tpot}.

This tells that the electroweak phase transition 
is of the first order. However, notice that 
the energy barrier is extremely small,
\begin{gather}
\frac{V_{eff}(\rho_-)
-V_{eff}(\rho_+)}{V_{eff}(\rho_+)} \Big|_{T_c}
\simeq 3.8 \times 10^{-6}.
\end{gather} 
Moreover, the barrier lasts only for short period since the temperature difference from $T_1$ to $T_c$ is very small, $\delta =(T_1-T_c)/T_c \simeq 0.0002$. So for all practical purposes we could treat the electroweak phase transition as a second order phase transition. 

\begin{figure}
\includegraphics[height=4cm, width=8cm]{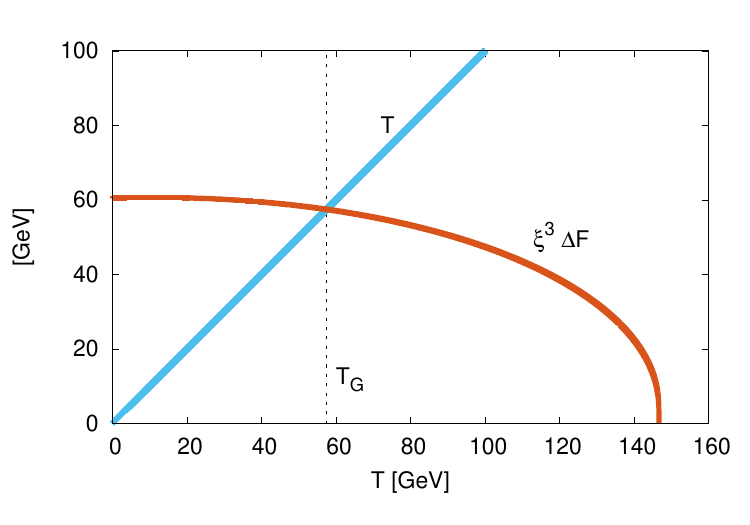}
\caption{\label{Gtemp} The determination of 
the Ginzburg temperature $T_G$ in the electroweak phase transition. Here the red and blue curve represents $\xi^3 \Delta F$ and the temperature of the universe, and the black line represents $T_G$.}
\end{figure}

The monopole production in the second order phase transition is supposed to be described by 
the Kibble-Zurek mechanism, so that the monopole production start from $T_c$. And the thermal fluctuations of the Higgs vacuum which create 
the seed of the monopoles continue as long as 
we have \cite{pta19,gin}
\begin{gather}
\xi^3 \Delta F \le T,
~~~~\Delta F(T)=V(\rho_s)-V(\rho_{+}),
\label{flcon}
\end{gather}
where $\xi(T)$ is the correlation length of the Higgs field and $\Delta F(T)$ is the difference in free energy density between two phases. This large fluctuation disappears when the equality holds, 
at the Ginzburg temperature $T_G$. 

We can find $T_G$ graphically from (\ref{epot}) and 
(\ref{flcon}). This is shown in Fig. \ref{Gtemp}. 
From this we have \cite{pta19}
\begin{gather}
T_G \simeq  57.49~\text{GeV},
~~~~\rho_+(T_G) \simeq 232.9~\text{GeV}.
\label{GT}
\end{gather} 
The effective potential at the Ginzburg temperature 
is shown in Fig. \ref{tpot}. Notice that $\rho_+(T_G)$ 
is quite close to the Higgs vacuum at zero 
temperature, which confirms that the electroweak monopole production lasts long time after the phase transition. 

With this observation we can say that the monopole formation takes place between $T_c$ and $T_G$, or 
roughly around $T_i$,
\begin{gather}
T_i= \frac{T_c+T_G}{2} \simeq  102.1~\text{GeV},   \nn\\
\rho_+(T_i) \simeq 188.4~\text{GeV}.
\label{itemp}
\end{gather}
To translate this in time scale, remember that the age 
of the universe $t$ in the radiation dominant era is 
given by \cite{kolb} 
\begin{gather}
t =\Big(\frac{90}{32 \pi G N_*(T)} \Big)^{1/2}~\frac{1}{T^2}.
\label{ewt}
\end{gather} 
So, with $N_* \simeq 385$ (including $\gamma, \nu, g, e, \mu, \pi, u, d, s, c, b, \tau, \\ W, Z, H$) we have 
\begin{gather}
t \simeq 0.048 \times \frac{M_P}{T^2}
\simeq 3.9 \times 10^{-7} \big(\frac{\rm GeV}{T} \big)^2 sec.	
\end{gather}  
From this we can say that the electroweak monopole 
production start from $1.8 \times 10^{-11} sec$ to 
$1.2 \times 10^{-10} sec$ after the big bang for 
the period of $10.3 \times 10^{-11}~sec$, or around 
$3.5 \times 10^{-11} sec$ after the big bang 
in average.

\begin{figure}
\includegraphics[height=7cm, width=8cm]{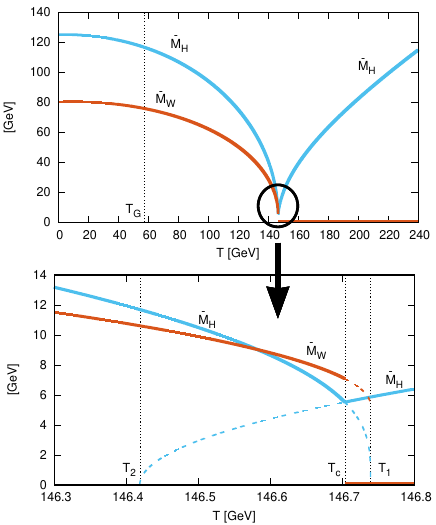}
\caption{\label{hwmass} The temperature dependent 
Higgs and W boson masses. The blue and red curves 
represent the Higgs and W boson masses.}
\end{figure}

The effective potential (\ref{epot}) gives us two 
important parameters of the electroweak phase transition, the temperature dependent Higgs mass 
$\bar M_H$ which determine the correlation length $\xi=1/ \bar M_H$, 
\begin{gather}
\bar M_H^2=\frac{d^2 V_{eff}}{d\rho^2} 
\Big|_{\rho_{min}}  \nn\\
=\left\{\begin{array}{ll} 
\Big[(\dfrac{T}{T_2})^2-1 \Big]
~\dfrac{M_H^2}{2}, &~~~T > T_c, \\ \Big[1+(\dfrac{\rho_+}{\rho_0})^2 -(\dfrac{T}{T_2})^2 \Big]~\dfrac{M_H^2}{2}, &~~~T \le T_c, 
\end{array} \right. 
\label{hmass}
\end{gather}
and the W-boson mass which determines the monopole 
mass $M \simeq \bar M_W / \alpha$, 
\begin{gather}
{\bar M}_W^2=\left\{\begin{array}{ll} 
0,&~~~T > T_c,  \\
\dfrac{g^2}{4} \rho_+^2,&~~~T \le T_c.	
\end{array} \right.  	
\label{wmass}
\end{gather}
The temperature dependent Higgs and W-boson masses 
are shown in Fig. \ref{hwmass}. 

Notice that the Higgs boson acquires the minimum 
mass $\bar M_H \simeq 5.5~{\rm GeV}$ at $T_c$ and approaches to the zero temperature mass 125.3 GeV as the universe cools down. Moreover, we have
\begin{gather}
\bar M_H (T_G) \simeq 116.9~{\rm GeV}.
\label{tdhm}
\end{gather} 
This confirms that at $T_G$ it already becomes close 
to the Higgs mass at zero temperature. But the W-boson 
which is massless above $T_c$ (before the symmetry breaking) becomes massive at $T_c$, and we have
\begin{gather}
\bar M_W (T_c) \simeq 7.1~{\rm GeV},	
~~~~\bar M_W (T_G) \simeq 76.0~{\rm GeV}.
\label{tdwm}
\end{gather} 
This implies that the infant monopole masses at $T_c$ 
and $T_G$ are (with $\bar M \simeq \bar M_W/\alpha$) 
around 1.0 TeV and 10.4 TeV (assuming the adolescent 
mass is $M_W/\alpha\simeq 11.0~\text{TeV}$). This 
tells that the baby monopole mass near $T_c$ could 
be considerably smaller than the adolescent monopole 
mass, although at $T_G$ the monopole mass becomes 
close to the adolecent value. This could allow LHC 
to produce the monopole even when the adolescent 
mass is bigger than 7 TeV. In Table I we list 
the masses of Higgs, W boson, and electrowak monopole at different temperatures for comparison.

\begin{table}
\begin{tabular}{|c||r|r|r|r|r|}
\hline
%&&&&& \\ 
\multicolumn{1}{|c|}{} &
\multicolumn{1}{|c|}{$T$} &
\multicolumn{1}{|c|}{$\rho_{+}(T)$} &
\multicolumn{1}{|c|}{$\bar M_H(T)$} &
\multicolumn{1}{|c|}{$\bar M_W(T)$} &
\multicolumn{1}{|c|}{$\bar M(T)$}
\\ \hline\hline
$T_1$ & \hspace{2mm} $146.74$ & $16.3$ & $5.9$ & $0$ & $0$
\\ \hline
$T_c$ & $146.70$ & $21.7$ & $5.5$ & $7.1$ & $971.7$
\\ \hline
$T_2$ & $146.42$ & $32.5$ & $11.7$ & $10.6$ & $1\,454.8$
\\ \hline
$T_i^{LHC}$ & $132.89$ & $119.2$ & $56.8$ & $38.9$ & $5\,331.5$
\\ \hline
$T_{min}^{LHC}$ & $119.36$ & $156.5$ & $76.2$ & $51.1$ & $7\,000.0$
\\ \hline
$T_i$ & $101.96$ & $188.4$ & $92.9$ & $61.5$ & $8\,428.2$
\\ \hline
$T_G$ & $57.49$ & $232.9$ & $116.9$ & $76.0$ & $10\,419.6$
\\ \hline
$0$ & $0.00$ & \hspace{5mm} $246.2$ & \hspace{5mm} $125.3$ & \hspace{5mm} $80.4$ & $11\,014.5$
\\ \hline
\end{tabular}
\caption{The values of $\rho_{+}$, $\bar M_H$, 
$\bar M_W$, and the expected monopole mass 
$\bar M =\bar M_W /\alpha$ at various temperatures. 
All numbers are in GeV.}
\label{TableValues}
\end{table}

This also confirms that what is important for 
the monopole production in the early universe is 
the Ginzburg temperature, not the type of the phase transition. The exponential suppression of 
the monopole production in the first order phase trabsition applies only when the Ginzburg 
temperature becomes higher than $T_2$. As far as 
the Ginzburg temperature becomes lower than $T_2$, there is no much difference between the monopole production in the first and second order phase transitions. 

\section{Topological Monopole Production by thermal fluctuation of Higgs vacuum at LHC}

If the fireballs made of the p-p (and heavy ion) collisions at LHC can reproduce the hot thermal 
bath of radiation of the early universe as it 
has often been asserted, we may assume that LHC produces the electroweak monopoles by the same mechanism that the early universe does. In this case, the above discussion suggests that LHC could produce the baby electroweak monopole of mass around 1.0 TeV just after $T_c$ and produce the last monopole of 
mass 10.4 TeV at the Ginzburg temperature $T_G$.    

However, the 14 TeV LHC can not produce the monopole 
of mass 10.4 TeV as it has to produce the monopole 
in pairs. This means that the present LHC can not produce the monopole exactly as the early universe does. In other words, although LHC could reproduce 
the early universe, it could do so only in a limited sense. This is because, unlike the big bang fueled 
by the infinite energy, LHC can provide only 
a finite energy. So we have to take care of this 
energy constraint at LHC. 

To do that, we estimate the temperature at which 
the monopole production stops at the present LHC. 
Since LHC should produce the monopole in pairs, 
the minimum temperature $T_{min}^{LHC}$ at the 14 
TeV LHC for the monopole production must satisfy 
the energy condition
\begin{gather}
M_{max} = 7~\text{TeV} \simeq \frac{\bar M_W(T_{min}^{LHC})}{\alpha} 
= \frac{g}{2} \frac{\rho_+(T_{min}^{LHC})}{\alpha}. 
\label{lhcec}
\end{gather}
Solving this we find 
\begin{gather}
T_{min}^{LHC} \simeq 119.4~{\rm GeV}.
\end{gather} 
This means that the energy condition to produce 
the monopole mass no more than 7 TeV forces 
the monopole production at LHC to stop at 119.4 GeV, much higher that the Ginzburg temperature. 
The effective potential at $T_{min}^{LHC}$ is also shown in Fig. \ref{tpot} in green line for comparison. 
  
If so, at LHC the electroweak monopole production 
starts at $T_c$ around 146.7 GeV and stops at $T_{min}^{LHC}$ around 119.4 GeV, not at the Ginzburg temperature around 57.5 GeV. In average the monopole production temperature at LHC is given by
\begin{gather}
T_i^{LHC}= \frac{T_c+T_{min}^{LHC}}{2} \simeq  132.9~\text{GeV},   \nn\\
\rho_+(T_i^{LHC}) \simeq 119.2~\text{GeV},
\label{lhct}
\end{gather}
not at $T_i$ given by (\ref{itemp}). This tells that 
LHC starts to produce the electroweak monopole with 
mass 1.0 TeV at around 146.7 GeV, and stops producing 
the monopole with mass 7 TeV at 119.4 TeV. In average 
LHC produces the infant electroweak monopole mass 
around 5.3 TeV, much less than the adolescent mass 
11.0 TeV. This implies that the 14 TeV LHC could actually produce the electroweak monopole pair even when the mass of the monopole pair becomes bigger 
than 14 TeV. This is remarkable. 

It is generally believed that in the Kibble-Zurek 
mechanism we are supposed to have one monopole per 
one correlation volume. This assumotion, however, 
may have a critical defect. This is because 
the correlation length is fixed by the electroweak scale but the monopole mass is given by $1/\alpha$ times bigger than the electroweak scale, so that 
the energy in one correlation volume may not be 
enough to make up the monopole mass. A natural way 
to cure this defect is to enlarge 
the correlation length $\xi =1/\bar M_H$ to 
$\bar \xi$,
\begin{gather}
\bar \xi=(\frac{1}{\alpha})^{1/3}~\xi
\simeq 5.16 \times \frac{1}{\bar M_H}.
\label{mcl}
\end{gather} 
With this we have the new correlation volume at 
$T_i^{LHC}$,
\begin{gather}
V_c \simeq \frac{4\pi^2}{3} \bar \xi^3
\simeq 0.76 \times 10^{-43} cm^3. 
\label{mcv}
\end{gather}
In comparison the p-p fireball volume at the 14 TeV 
LHC is given by (with the proton radius $0.87 \times 10^{-13}~cm$) 
\begin{gather}
V_{pp} \simeq \frac{0.938}{7000} \times \frac{4\pi^2}{3} 
~(0.87)^3 \times 10^{-39} cm^3  \nn\\
\simeq 1.16 \times 10^{-42} cm^3,
\label{ppv}	
\end{gather}	   
where the first term represents the Lorentz contraction
of the p-p fireball. Notice that $V_c$ is about thirteen times bigger than $V_{pp}$. This tells that the p-p fireball at LHC has just enough size to produce the monopole.     
  
We could translate the monopole production process 
at LHC in time scale. From (\ref{ewt}) we can say 
that the electroweak monopole production at LHC 
starts at 146.7 GeV for the period of 
$0.7 \times 10^{-11}~sec$. This tells that 
the monopole production lasts only very short period 
at LHC. So one might wonder if we have enough 
thermal fluctuations of the Higgs vacuum during 
this period. We can estimate how many times the Higgs 
vacuum fluctuates from $\rho_+(T_{LHC})$ to zero in 
average. From the uncertainty principle the time 
$\Delta t$ for one fluctuation is given by
\begin{gather}
\Delta t \simeq \frac{1}{\Delta E}
\simeq 4.7 \times 10^{-27}~sec,	
\end{gather} 
so that the number of the fluctuation $N_f$ of 
the Higgs vacuum is given by
\begin{gather}
N_f \simeq \frac{\bar t}{\Delta t}
\simeq 3.1 \times 10^{16}.	
\end{gather} 
This assures that we have enough fluctuations of 
the Higgs vacuum to produce the monopoles at LHC.

The above discussion tells that LHC could produce 
the electroweak baby monopole of mass around 
5.3 TeV even when the adolescent mass becomes 
11.0 TeV, if we assume that the monopole production mechanism at LHC is thermal. If this is true,
this would be really remarkable. 

Unfortunately, the above argument has a critical 
defect, because here we have implicitly assumed 
that the temperature of the p-p fireball would be 
high enough to produce the monopole thermally 
(assuming that the 14 TeV p-p fireball energy 
would produce the high temperature necessary for 
LHC to produce the monopole thermally). However, 
this is not obvious, because the temperature and 
energy of the fireball are two different things. 
As we have argued, for LHC to produce the monopole 
thermally, the temperature of the fireball must 
be no less than 119.4 GeV. And if the temperature 
of the fireball higher than this, the above 
conclusion becomes valid. But is this really so? 
This is a non-trivial question. 

A straightforward way to answer this question is to measure the temperature of the p-p firebal at LHC.
Fortunately the ALICE group has actually measured 
the temperature of the p-p fireball of the center 
of mass energy 2.76 TeV at LHC, and found that 
the temperature is around 297 MeV, which is roughly $10^{-4}$ times smaller than the center of mass 
energy of the p-p fireball \cite{alicet}. The ALICE result was unexpected, but clearly shows that 
the temperature of LHC p-p fireball is much less 
than the temperature needed to produce the monopole 
thermally, even with the future FCC energy.  

This tells that here again, the thermal production 
of the electroweak monopole at LHC is practically impossible, although it is logically possible. 
This teaches us an important lesson. If the monopole production mechanism at LHC is the thermal production, it is virtually impossible for LHC to produce 
the monopole, simply because the temperature of 
the p-p fireball is too low. In this case the only 
way to detect the electroweak monopole is to look 
for the remnant monopoles produced in the earluy universe.
 
This also seems to suggest that the Drell-Yan process and/or the Schwinger mechanism might be the only probable way for LHC to produce the monopole.
Obviously, the present 14 TeV LHC could produce 
the monopole when the mass becomes less than 7 TeV, 
if the monopole production mechanism becomes 
the Drell-Yan process and/or the Schwinger mechanism. 

On the other hand, there may be another logical possibility for LHC to produce the monopole, 
the monopole production in the form of the monopolium bound states, as we have pointed out. An important feature of this mechanism is that the binding energy 
of the monopolium could reduce the mass below 14 TeV even when the mass of the monopole-antimonopole 
pair becomes 22 TeV. In this case LHC could produce 
the monopolium even when the monopole mass becomes 
11 TeV or heavier. This necessiate us to study 
the monopolium bound state in more detail, which we 
discuss in the following.

\section{Electroweak Monopolium: Atomic Model}

The monopolium itself has been studied before for various reasons \cite{nambu,hill,epele}. For example, Nambu proposed to view the monopolium as a model of strong interaction, because the magnetic coupling 
of the monopole-antionopole pair drastically 
increases the energy spectrum of the monopole pair 
to the order of hundred MeV \cite{nambu}. But here 
we are interested in the electroweak monopolium 
made of the Cho-Maison monopole pair. 

To discuss the electroweak monopolium we let 
the monopole mass be $M$ and consider 
the Schr\"odinger equation of the monopolium wave function 
\begin{gather}
\Big(-\frac{1}{2\mu} \nabla^2 + V \Big) \Psi= E~\Psi, \nn\\
V= -\frac{4\pi}{e^2} \frac{1}{r},
\label{seq}	
\end{gather}
where $\mu=M/2$ is the reduced mass and $V$ is 
the Coulombic magnetic potential of 
the monopole-antimonopole pair. Obviously this is 
formally identical to the Schr\"odinger equation of 
the Hydrogen atom, except that the coupling strenth 
of the potential is replaced by $4\pi/e$ and 
the electron mass is replaced by $\mu$. 

From this we have the energy spectrum 
\begin{gather}
E_n = -\frac{M}{4 \alpha^2 n^2} 
= -\frac{4,692.3}{n^2} \times M  \nn\\
\simeq -\frac{51,615.3}{n^2}~{\rm TeV}, 	
\label{aen}
\end{gather}  
where $\alpha=e^2/4\pi$ is the fine structure constant 
and we have put $M=M_W/\alpha \simeq 11.0~\text{TeV}$. 
So the Rydburg energy of the monopolium becomes 
51,615 TeV. Moreover, the Bohr radius $R$ of 
the monopolium is given by 
\begin{gather}
R = \frac{\alpha}{\mu} \times \alpha^2
= \frac{2 \alpha^3}{M} 
\simeq \frac{5.6\times~10^{-9}}{M_W} \nn\\
\simeq 1.4 \times 10^{-24} cm,
\label{br}
\end{gather}  
Notice, however, that these results are totally 
unreasonable. First, the above Rydburg energy is 
too big (roughly 2,346 times bigger than 
the monopole-antimonopole mass), which means that 
the monopole mass simply can not provide the binding energy necessary to make the monopolium. Second, 
the monopolium size (i.e., the Bohr radius) is $10^{-8}$ times smaller than the monopole size,
which is absurd.

\begin{figure}
\includegraphics[height=3cm,  width=7cm]{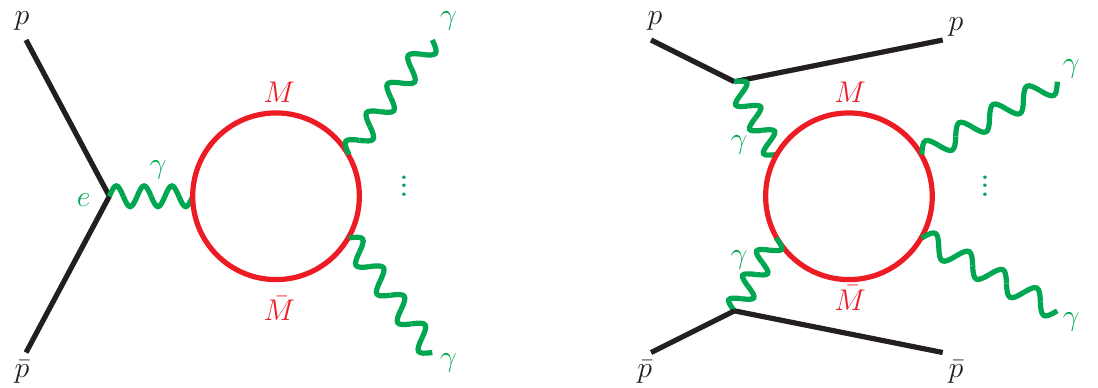}
\caption{\label{lhcmpm} The monopole production 
mechanismin the form of the monopolium at LHC.}
\end{figure}

This could mean the following. First, the monopolium 
quickly annihilates itself, long before it decays to 
the ground state. In fact (\ref{aen}) suggests that 
the monopolium could exist only at highly excited 
levels, probably with $n \geq 49$, because below 
this level the monopole-antimonopole mass can not provide the necessary binding energy. Or else, this could mean that the naive atomic description of 
the monopolium is not acceptable. In this case we 
have to find a more realistic model of the monopolium.

The reason for the above results originates from 
the fact that the magnetic coupling of the monopole 
is too strong, which makes the monopolium size 
too small. We could make the atomic model of 
the monopolium more realistic by improving 
on this point. For example, we could modify 
the magnetic Coulombic potential at short distance 
for $r \leq 1/M_W$, to accommodate the fact that 
the monopole has a finite size $1/M_W$. 

On the other hand the above exercise has one positive 
side, because it implies that the binding energy of 
the electroweak monopolium could be big, of the order 
of TeV. This could make the monopolium mass considerably less than $2M$, so that we could have 
the monopolium of mass lighter than 14 TeV even when the mass of monopole-antimonopole pair is heavier 
than 14 TeV, with a more realistic monopolium potential. 

\section{Electroweak Monopole Production at LHC via Monopolium}

Now, we are ready to discuss the new monopole production mechanism in the form of the monopolium,
which could allow us to prove the existence of 
the electroweak monopole at LHC even when 
the monopole mass exceeds 11 TeV. This is 
schimatically shown in Fig. \ref{lhcmpm}. Notice that
LHC could produce the monopole with this mechanism,
when the monopole production mechanism at LHC becomes
the Drell-Yan process and/or the Schwinger mechanism
but the energy condition prevent the production of 
the monopole when the monopole mass becomes bigger 
than 7 TeV. 

In fact the first stage of this mechanism is 
exactly like the Drell-Yan process which produces 
the monopole-antimonopole pair. But when the monopole mass exceeds 7 TeV, the pair can not be materialized and could exist only vertually. So, the Drell-Yan process becomes ``virtual", and the final state of 
this mechanism becomes not the monopole-antimonopole pair but the decay modes of the monopolium. 

We could have a similar situation with the Schwinger mechanism. Here again the energy condition of LHC 
could force the monopole-antimonopole pair virtual, 
and the final state could become the decay modes 
of the monopolium. In this sence this monopole production mechanism could be thought as the virtual Drell-Yan process or the virtual Schwinger mechanism. 
On the other hand, we emphasize that the final 
state of this monopole production mechnism is 
totally different from the Drell-Yan process 
and/or the Schwinger mechanism.

To show that this mechanism could indeed prove
the existence of the electroweak monopole at LHC 
even when the monopole mass exceeds 11 TeV, we must have a realistic model of the monopolium whose 
binding enerfy could reduce the monopolum mass 
below 14 TeV. For this we have to remember that intuitively the size of the electroweak monopolium 
can not be smaller than $2/M_W$, considering 
the fact that the monopole has the size of $1/M_W$. This means that at short distance the singularity 
of the magnetic potential in (\ref{seq}) should be regularized. 

\begin{figure}
\includegraphics[height= 4.5cm, width=7.5cm]{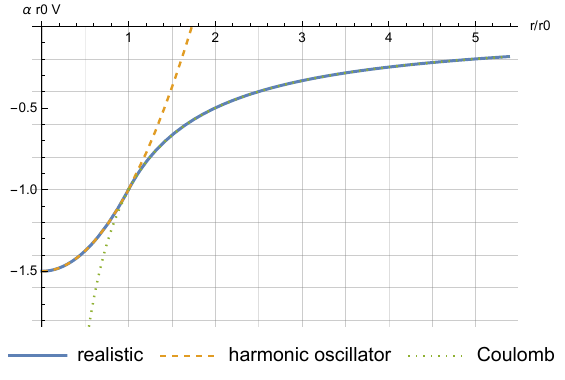}
\caption{\small The modified magnetic potential (\ref{mp}) (multiplied by $r_0$) of the electroweak monopolium shown as a function of $r/r_0$.}
\label{mpot}
\end{figure}

So we modify the potential $V$ at short distance 
with a harmonic oscillator potential smoothly 
connected at $r_0$, and let
\begin{gather}
V =\left \{\begin{array}{rcl} 
\dfrac{2\pi}{e^2 r_0} \Big(\dfrac{r^2}{r_0^2}-3 \Big),
&~~~r< r_0, \\~\\
-\dfrac{4\pi}{e^2} \dfrac{1}{r},~~~~~~~~&
~~~r \geq r_0, 
\end{array} \right.
\label{mp}
\end{gather}
where $r_0$ is of the order of $1/M_W$. The potential is shown in Fig. \ref{mpot} in the unit of $r_0$. 

The significance of this potential is that 
the coupling constant $k=2\pi/e^2r_0$ of the hamornic oscillator potential is automatically fixed by 
the smoothness of the potential at $r_0$. This immediately tells that the energy gap of the lowlying energy spectrum is given (with $r_0 \simeq 1/M_W$) roughly by $M_W/2\alpha=M/2$, which reduces the energy gap of (\ref{aen}) by the factor $10^{-3}$. This implies that the modified potential (\ref{mp}) could describe a realistic electroweak monopolium. With 
this observation we could solve the Schr\"odinger 
equation keeping the reduced mass $\mu$ of 
the monopole-antimonopole pair arbitrary, allowing 
$\mu$ to have a small mass.

The first step to do this is the usual separation of variables of the wave function in the regions $r<r_0$ (the region I) and $r>r_0$ (the region II) as
\begin{gather}
\Psi_{\rm{I,II}}(r,\theta,\varphi) \equiv \frac{R_{\rm{I,II}}(r)}{r}Y_{l}^{m}(\theta,\varphi), 
\end{gather}
where $Y_{l}^{m}(\theta,\varphi)$ are the spherical 
harmonics. Next, we adopt dimensionless coordinates 
\begin{gather}
r = \frac{2\alpha}{M}x\,~~~~r_0 = \frac{2\alpha}{M}x_0,
~~~~E \equiv -\frac{M}{4\alpha^2}\epsilon,
\end{gather}
where we denoted dimensionless binding energy as 
$\epsilon >0 $. In these coordinates, the relevant Schr\"odinger equations for the radial parts read
\begin{gather}
\ddot R_{\rm I} -\Big(\frac{l(l+1)}{x^2}
+\frac{x^2}{x_0^3} -\frac{3}{x_0}
+\epsilon \Big) R_{\rm I} =  0, \nn\\
\ddot R_{\rm II} -\Big(\frac{l(l+1)}{x^2}-\frac{2}{x_0}
+\epsilon\Big) R_{\rm II} =  0. 
\end{gather}
Both of these can be mapped to the Kummer's equation,
\begin{gather}	
z \frac{d^2 w(z)}{d z^2} 
+(b-z)\frac{d w(z)}{d z}-a w(z) =0,
\end{gather}
by suitable choices of $z \equiv z (x)$ and parameters
$a$ and $b$. 

In particular, this mapping is facilitated via assignments,
\begin{gather}
R_{\rm I} =x^{l+1} \exp \Big(-\frac{x^2}{2x_0^{3/2}} \Big) w(x^2/x_0^{3/2}), \nn\\
a_{\rm I} =\frac{2l+3+x_0^{3/2}\epsilon
-3\sqrt{x_0}}{4}, 
~~~~b_{\rm I} = l+3/2, 
\end{gather}
and 
\begin{gather}
R_{\rm II} = x^{l+1} \exp \big(-\sqrt{\epsilon} x \big) w(2\sqrt{\epsilon}x), \nn\\
a_{\rm II} = l+1 -\epsilon^{-1/2},
~~~b_{\rm II} = 2l+2.
\end{gather}
As is well-known, the Kummer's equation has two 
independent solutions, the confluent hypergeometric function $M(a,b,z) \equiv\, _1 F_1\bigl(\begin{smallmatrix}a \\ 
b\end{smallmatrix}\bigr| z\bigr)$ and the Tricomi's function $U(a,b,z)$. 

In the region I ($x<x_0$), we need to satisfy 
the regularity condition at the origin $x=0$, which 
forces us to exclude $U(a,b,z)$ as it has a singularity 
at $z=0$ in contrast to $M(a,b,z)$, which is an entire 
function in $z$. Hence, the general solution in 
the first region that is regular at the origin is 
given by 
\begin{gather}
R_{\rm I} = N_{\rm I}\, x^{l+1} 
\exp \Big(-\frac{x^2}{2x_0^{3/2}} \Big)  \nn\\
\times _1F_1\Big(\begin{matrix}\frac{l}{2}
+\frac{3}{4}+\frac{x_0^{3/2}\epsilon-3 \sqrt{x_0}}{4} \\ l+\frac{3}{2}\end{matrix}\Bigr|\tfrac{x^2}{x_0^{3/2}}\Big),
\label{r1}
\end{gather}
where $N_{\rm I}$ is an arbitrary constant at 
the moment which should be determined later. 

In the region II ($x>x_0$), we need to impose 
boundary condition at $x\to \infty$ so that 
the resulting wave equation is square-integrable. 
This forces us to discard $M(a,b,z)$ that behaves as 
$M(a,b,z) \sim \Gamma(b) \exp (z)~z^{a-b}/\Gamma(a)$ 
for large $z$ (in a certain wedge of the complex plane). Notice that this asymptotic behaviour holds only if $a$ is not a negative integer (and hence $\Gamma(a)$ blows up), which is exactly the case 
for the hydrogen atom. For our modified potential, however, the spectrum differs and hence $a \neq -k$, where $k$ is a positive integer. 

On the other hand, we have
\begin{gather}
U(a,b,x) \sim x^{-a}\, _2 F_0 \bigl(\begin{matrix}a 
& a-b+1 \\ & \end{matrix}\bigr| -1/x\bigr),
\end{gather}
as $x$ goes to infinity. Hence, the general solution 
in the region II has the following form,
\begin{gather}
R_{\rm II} = N_{\rm II}\, x^{l+1} 
\exp \big(-\sqrt{\epsilon}x \big)   \nn\\
\times U\Big(\begin{matrix}l+1 -\frac{1}{\sqrt{\epsilon}} \\ 2l+2\end{matrix}\Bigr| 2\sqrt{\epsilon}x\Big),
\label{r2}
\end{gather}
where $N_{\rm II}$ is again an arbitrary constant
for the moment. 

The remaining task is to formulate sewing condition that ensures $C_1$ continuity of the wavefunction at the point $x=x_0$. The condition that $R_{\rm I}(x_0) =R_{\rm II}(x_0)$ fixes $N_{\rm II}$ in terms of $N_{\rm I}$, which can be in turn fixed by square integrability condition. The other condition 
$R_{\rm I}^\prime(x_0) = R_{\rm II}^\prime(x_0)$ 
then boils down to the following transcendental equation for the binding energy,
\begin{gather}
\frac{3+2l-3x_0^{1/2}+x_0^{3/2}\epsilon}{3+2l}
\frac{_1 F_1 \Big(\begin{matrix}\frac{7+2l-3x_0^{1/2}
+x_0^{3/2}\epsilon}{4} \\ 
\frac{5}{2}+l\end{matrix}\Bigr| x_0^{1/2}\Big)}
{_1 F_1 \Big(\begin{matrix}\frac{3+2l-3x_0^{1/2}
+x_0^{3/2}\epsilon}{4} \\ 
\frac{3}{2}+l\end{matrix}\Bigr| x_0^{1/2}\Big)}
\nn\\
+2\sqrt{x_0}\big((1+l)\sqrt{\epsilon}-1\big)
\frac{U \Big(\begin{matrix}2+l-\epsilon^{-1/2} \\ 3+2l\end{matrix}\Big| 2x_0 \epsilon^{1/2}\Big)}
{U \Big(\begin{matrix}1+l-\epsilon^{-1/2} \\ 2+2l
\end{matrix}\Bigr| 2x_0 \epsilon^{1/2}\Big)} \nn\\ 
+\sqrt{\epsilon x_0} = 1.
\label{cond1}
\end{gather} 
With the two conditions we can obtain the smooth 
radial wave function and the energy spectrum of 
the monopolium. 

\begin{figure}
\includegraphics[height=5.5cm,
width=\linewidth]{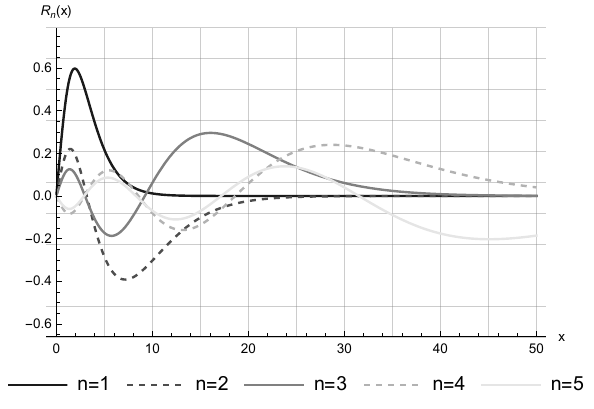}
\caption{The radial monopolium wave function R(x) 
for $n=1,2,3,4,5$ shown in (\ref{r1}), (\ref{r2}), 
and (\ref{cond1}).} 
\label{mwf}
\end{figure}

\begin{figure}
\includegraphics[height=5.5cm,
width=\linewidth]{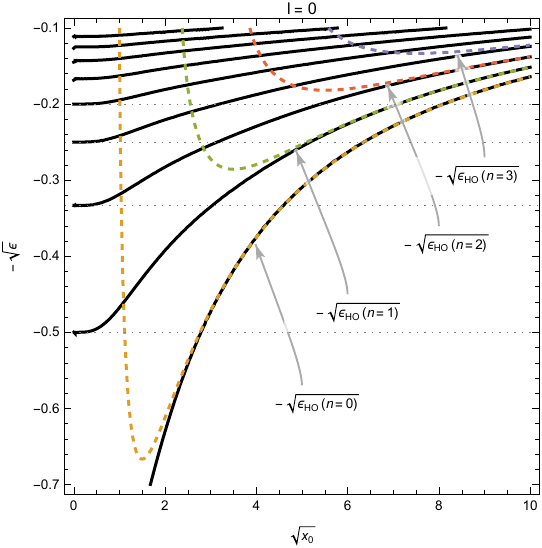}
\caption{The energy spectrum $\sqrt{\epsilon}$ of 
the monopolium with $l=0$ fixed by (\ref{cond1}), 
as function of $\sqrt{x_0}$. The horizontal dotted lines are the energy levels of the hydrogen atom 
given by (\ref{ae}), and the dashed lines represent harmonic oscillator energy levels given by (\ref{harmosc}).}
\label{cond} 
\end{figure}

The radial wave function for $n=1,2,...,5$ is shown 
in Fig. \ref{mwf}. However, the condition (\ref{cond1}) which determines the energy spectrum of the monopolium is not easy to solve, and it can only be solved numerically. Fortunately we could solve it, and 
the resulting energy spectrum is displayed in 
Fig. \ref{cond} (for $l=0$). 

We see that as $x_0$ goes to zero, the binding energies approaches the values given by the atomic model
\begin{gather}
\epsilon_A \simeq \frac{1}{n^2},~~~~n=1, 2,\ldots,
\label{ae}
\end{gather}
whereas for large $x_0$, the contours quickly settles on the curves given by the  binding energies of 
the harmonic oscillator, i.e.,
\begin{gather}
\epsilon_{HO} = \frac{3\sqrt{x_0}-3-2l-4n}{x_0^{3/2}}, 
~~~~n = 0, 1, \ldots 	
\label{harmosc}
\end{gather}
The value of $x_0$ we are interested in is given by  
$r_0 \sim 1/M_W$, or
\begin{gather}
x_0 = \frac{M}{2\alpha} r_0 \simeq 7,767,173.
\end{gather}
For such a high values of $x_0$, the binding energies are almost indistinguishable from the binding energies of harmonic oscillator. 

This means that the ground state energy of 
the monopolium (for $l=0$) is roughly 
\begin{gather}
E_0 = -\frac{\epsilon_0}{4\alpha^2}~M
\simeq -4,692 \times (M/\text{TeV})~\epsilon_{HO}^{n,l=0} \nn\\
\simeq -16.8~\text{TeV}.
\label{maen}
\end{gather}
where we have put $M\simeq 11.0~\text{TeV}$. Compared 
with (\ref{aen}), this is roughly $3 \times 10^{-4}$ 
times reduction of the naive estimate of the binding 
energy given by hydrogen-like atomic model of 
the monopolium. In particular, this tells that 
the monopole-antimonopole pair with mass 22 TeV 
makes the monopolium bound state of mass of 5.2 TeV 
at the ground state. This is precisely what we have 
hoped for.  

Furthermore, the Bohr radius of the monopolium, which 
can be read of from the exponential decay of $R_{\rm II}$ from (\ref{r2}) is given by
\begin{gather}
R = \sqrt{\frac{1}{|E_0|}} \simeq 1.5 \times 10^{-18}~cm.
\label{mbr}
\end{gather}
This is $0.6 \times 10^{-2}$ times smaller 
than the monopole size given by 
$1/M_W \simeq 2.5 \times 10^{-16}~cm$.  Again, this 
is definitely much more realistic than (\ref{br}). 
This confirms that the monopolium bound state with 
the modified potential (\ref{mpot}) can give us 
the desired binding energy which could help LHC to produce the bound state even when the monopole mass becomes bigger than 7 TeV.  

Notice that in the above analysis we have assumed 
the monopole mass to be 11 TeV to have the monopolium mass 5.2 TeV. And (\ref{maen}) tells that 
the monopolium mass becomes 3.3 TeV when the monopole mass becomes 7 TeV. Perhaps more importantly, this tells that the maximum mass of the monopole that 
the present 14 TeV LHC can produce the monopolium of mass 14 TeV becomes 29.8 TeV. This strongly implies that the present LHC could produce the electroweak monopole in the form of the monopolium bound states
for any reasonable monopole mass (as far as the mass remains less than 29.8 TeV), if the monopole production mechanisn becomes the Drell-Yan process.

Of course, the monopolium mass at LHC may not 
turn out to be the same as the above prediction, because we do not know how realistic our potential
(\ref{mp}) is. In this connection it should be mentioned that a different model of the monopolium which assumes an exponential repulsion at short distance given by the potential
\begin{gather}
V =-\frac{4\pi}{e^2} \frac{1}{r} 
\Big(1-\exp(-r/2r_0) \Big),
\end{gather}
has been discussed before \cite{epele}. Despite  
the obvious difference between our potential 
(\ref{mp}) and this, we notice that the characteristic features of the two monopoliums are quite similar. 
In particular, in both cases the monopolium masses 
are slightly below the half of the monopole mass 
and the radius are of the order of $10^{-18} cm$. 
This implies that our results discussed above are reliable.

\section{Discussion}

An urgent task at LHC is to discover the electroweak monopole which exists within the standard model. Because of the unique charactistics of the electroweak monopole, the MoEDAL detector may have no difficulty 
to detect the monopole, if LHC could produce it. But the problem is that intuitively the present 14 TeV 
LHC may not be able to produce the monopole, if 
the monopole mass becomes bigger than 7 TeV. And indeed, if we suppose the monopole mass is $1/\alpha$ times the W boson mass, the present LHC cannot 
produce it. In this case the detection of the monopole at LHC may become hopeless, and we may have to look 
for the remnant electroweak monopoles which could 
have been produced in the early universe after 
the electroweak phase transition \cite{pta19,cko}.  

In this paper we have shown that the question if 
the present LHC could produce the monopole or not
critically depends on the monopole production 
mechanism at LHC. Our work in this paper strongly implies the followings. First, if the monopole production mechanism at LHC is the topological thermal 
fluctuation of the Higgs vacuum, there is practically no hope that it could produce the monopole, simply because the temperature of the p-p fireball is too 
low to produce the monopole thermally. And this 
would be the case even with the new FCC. This is unexpected, but if the recent ALICE measurement 
of the temperature of the p-p fireball is trustable, this conclusion is unavoidable.     

Second, if the monopole production mechanism at LHC 
is the Drell-Yan process, the present 14 TeV LHC 
could produce the monopole only if the monopole 
mass becomes less than 7 TeV. This was expected. 

In this paper we discussed a new monopole production mechanism, the monopole production in the form of 
the monopolium bound state, which could allow LHC 
to produce the electroweak monopole for any reasonable mass of the monopole, as far as the mass does not exceed 29,7 TeV. This is because the monopolium 
binding energy greatly reduces the mass of 
the monopolium. This tells that the present LHC 
could circumbent the energy constraint and produce 
the electroweak monopole in the form of the monopolium bound state, even when the monopole mass becomes 
bigger than 7 TeV. In this case we MoEDAL, ATLAS, 
and CMS could detect the decay modes of the monopolium and thus confirm the existence of the electroweak 
monopole at LHC. This is remarkable.

Of course, with the Schwinger mechanism at LHC 
we can have the same conclusion, if the p-p fireball could create strong enough magnetic background 
which can actually produce the monopole-antimonopole pairs. But it is by no means clear how the p-p 
fireball at LHC could generate such a strong 
magnetic background.

To summarize, our results tells that a most probable way for LHC to produce the monopole is the monopole
production in the form of the monopolium, and the 14 TeV LHC could more likely produce the monopole in the form of the monopolium bound states rather than as individual monopoles. In this case the most realistic way to detect the monopole at LHC is to detect 
the two photon decay modes of the monopoliums. And 
in principle MoEDAL, ATLAS, and CMS could do that. 

Finally, it should also be emphasized that there 
is the possibility that the LHC may not produce 
the monopole signal at all. This is because neither 
the Drell-Yan process nor the Schwinger mechanism
could turn out to be the monopole production 
mechanism at LHC. In this case the only way to 
detect the electroweak monopole could be to look 
for the remnant monopoles produced during 
the early universe, with the ``cosmic" MoEDAL. 
This makes the detection of the remnant monopoles produced during the early universe very important. 
 
{\bf ACKNOWLEDGEMENT}

~~~PB and FB are supported by the Institute of Experimental and Applied Physics, Czech Technical University in Prague and the Research Centre for Theoretical Physics and Astrophysics, Institute of Physics, Silesian University in Opava. YMC is 
supported in part by the National Research Foundation of Korea funded by the Ministry of Science and Technology (Grant 2022-R1A2C1006999 and 
2025-R1A2C17064731) and by Center for Quantum Spacetime, Sogang University, Korea.

\end{document}